%% file: main.tex
\documentclass[letterpaper, 10 pt, conference]{ieeeconf}  

\IEEEoverridecommandlockouts                              
\usepackage[utf8]{inputenc}
\usepackage[T1]{fontenc}

\usepackage{amsmath,cite,url}
\usepackage{graphicx}
\usepackage{color}
\usepackage{graphics} 
\usepackage{epsfig} 
\usepackage{mathptmx} 
\usepackage{amsmath} 
\usepackage{amssymb}  
\usepackage{xcolor}

\usepackage{multirow}
\usepackage{subcaption}
\usepackage{threeparttable}

\title{\LARGE \bf
 Receptive-Field  Regularized CNNs for Music Classification and Tagging
}

\author{Khaled Koutini,  Hamid Eghbal-Zadeh, Verena Haunschmid, Paul Primus,  Shreyan Chowdhury,  Gerhard Widmer \\ Johannes Kepler University Linz\\
firstname.lastname@jku.at}

\begin{document}

\maketitle
%

\begin{abstract}
Convolutional Neural Networks (CNNs) have been successfully used
in various Music Information Retrieval (MIR) tasks, both as end-to-end models and as feature extractors
for more complex systems. However, the MIR field is still dominated by the
classical VGG-based CNN architecture variants, often in combination with more complex modules such as attention, and/or techniques such as pre-training on large datasets.
Deeper models such as ResNet -- which surpassed VGG by a large margin in other domains -- are rarely
used in MIR. 
One of the main reasons for this, as we will show, is the lack of
generalization of deeper CNNs in the music domain.

In this paper, we present a principled way to make deep architectures like
ResNet competitive for music-related tasks, based on
well-designed regularization strategies. In particular, we analyze the
recently introduced \textit{Receptive-Field Regularization} and \textit{Shake-Shake},
and show that they significantly improve the generalization of deep CNNs on
music-related tasks, and that the resulting deep CNNs can
outperform current more complex models such as CNNs augmented with pre-training and attention.
We demonstrate this on two different MIR tasks and two corresponding
datasets, thus offering our deep regularized CNNs as a new baseline for
these datasets, which can also be used as a feature-extracting
module in future, more complex approaches.

\end{abstract}

\section{Introduction and Related work}\label{sec:introduction}
\input{s1_intro}

\section{Regularization approaches }\label{sec:regularizers}
\input{s2_regular}

\section{Musical Tasks}\label{sec:tasks}
\input{s3_tasks}

\section{Architectures}\label{sec:architectures}
\input{s4_archs}

\section{Experimental Setup}\label{sec:setup}
\input{s5_setup}

\section{Results}\label{sec:results}
\input{s6_results}

\section{Conclusion}\label{sec:conclusion}
\input{s7_conclusion}

\section{Acknowledgments}

This work has been supported by the LCM – K2 Center within the framework of the Austrian COMET-K2 program,
and the European Research Council (ERC) under the EU's Horizon 2020 research and innovation programme, under grant agreement No 670035 (project ``Con Espressione'').

\bibliography{refs}
\bibliographystyle{ieeetr}
%
%
%
%

\end{document}

%% file: s1_intro.tex
Convolutional Neural Networks (CNNs) have been successfully used
in various Music Information Retrieval (MIR) tasks, both as
end-to-end models and as feature extractors for more complex systems.
For instance -- to take just the two MIR problems that we will focus on in this
paper --, CNNs have become an essential tool for tasks like mono- and polyphonic
instrument recognition \cite{LiQW15Instrument,GururaniSL18instrument,HanKL17Instrument,GururaniSL19openmoc_attn,humphrey2018openmic}
or general music auto-tagging from spectrograms~\cite{choi2016automatic} or waveforms~\cite{pons2017end}.

In recent years, deep learning research has been progressing in various directions
such as designing new architecture topologies,  building specialised processing units,
and developing transfer learning techniques.
For example, in machine vision, very deep architectures such as
ResNet~\cite{heDeepResidualLearning2016} and
DenseNet~\cite{huangDenselyConnectedConvolutional2017} have been shown
to yield superior performance compared to VGG~\cite{simonyanVeryDeepConvolutional2014}
architectures, due to the special topological structure of their convolutional layers.
Also, the benefit of transfer learning, in the form of pre-training models on large
datasets such as ImageNet~\cite{deng2009imagenet}, has been demonstrated on various
vision tasks~\cite{kornblith2019better}.
In the field of NLP, attention mechanisms~\cite{vaswani2017attention} have been proposed
as a new processing unit, and have established a new state-of-the-art.

Following this trend, the MIR community has adopted many of these success stories
for various music-related tasks.
In instrument recognition, for instance, some recent classification
methods make use of \textit{pre-trained} feature-extraction CNNs to obtain low-dimensional
spectrogram embeddings~\cite{hersheyCNNArchitecturesLargescale2017,humphrey2018openmic};
these feature extraction CNNs are typically trained on large-scale datasets, such as
AudioSet~\cite{audioset2017Gemmeke}.
Moreover, \cite{GururaniSL19openmoc_attn} and \cite{ amir2020openmic} incorporate an
\textit{attention mechanism}, in combination with embeddings of a pre-trained CNN.
In music auto-tagging, recent work employing self-attention with CNN front-end and
CNN or CRNN (convolutional RNN) back-end have achieved competitive results on the
tagging task, with a focus on interpretability~\cite{won2019toward}. There have also
been successful attempts at using musically-motivated CNNs~\cite{pons2016experimenting}
that tune the convolution filter sizes in the time or frequency directions. 

However, very deep architectures such as ResNet, which have
proven so successful in domains like machine vision, have met with limited
success in the MIR field.
In this paper, we shed light on why such convolutional architectures
do not perform well on some MIR tasks and datasets, relating the problem
to their generalization ability, which again is due to their depth
and the consequent growth of their receptive fields (RFs).
In particular, we will show how the generalization ability of CNN models is
related to their RFs. 
Finally, we analyze two regularization techniques -- Receptive Field (RF) Regularization
and Shake-Shake Regularization -- for CNNs,
and show that with such simple regularization techniques, fully convolutional models
without any attention or pre-training on large datasets, can outperform more complex approaches.
For demonstration and analysis, we choose two specific MIR tasks -- emotion and theme auto-tagging
and instrument recognition -- along with pertinent benchmark datasets.
We also release our trained models and experiments source code\footnote{\url{https://github.com/kkoutini/cpjku_dcase19}},
to offer our models to the MIR community as a new baseline for these datasets. 

%% file: s2_regular.tex
In this section, we introduce the regularization techniques that we investigate in our experiments. We study two techniques: (a) Receptive-Field Regularization and (b) Shake-Shake Regularization. Both of these can be seen as a kind of inductive bias introduced in the architectural design of the CNN itself, regardless of the learning and optimization regulators.
\subsection{Receptive-Field Regularization}
Receptive-Field (RF) Regularization was introduced by Koutini et al.~\cite{Koutini2019Receptive} in the context of acoustic scene classification tasks.
They show that CNNs with larger receptive fields over the input spectograms tend to overfit on the training scenes and 
 have a lower performance when tested on new scenes and environments. In other words,
CNNs with larger RF -- such as common deep CNNs used in the vision domain -- learn features from the spectograms that don't generalize well. CNNs that utilize this regularization technique have shown a lot of success in different acoustic tasks and achieved top ranks in challenges and benchmarks, in particular, device-invariant acoustic scene classification~\cite{Mesaros2019,Primus2019,Koutinitrrfcnns2019,Suh2020task1a,Koutini2020dcasesubmission}, audio tagging with noisy labels and minimal supervision~\cite{Fonseca2019,Koutinitrrfcnns2019}, and open set acoustic scene classification~\cite{Mesaros2019,Lehner2019}, low-complexity acoustic scene classification. 

The RF of a neuron in a convolutional layer refers to the slice of the layer input that influences the neuron's activation.  A convolutional layer input can be the activations of a previous convolution layer or the network input; in both cases, we call the layer's input feature map. The size of the RF of each neuron w.r.t the layer's input feature map is determined by the size of its filter: the larger the filter, the larger the portion of the input feature map that can affect the neuron's activation. The receptive field $RF_n$  of neurons in a layer $n$  w.r.t the \textit{input spectograms} is affected by all the previous layers, and can be calculated recursively as shown in Equation~\ref{eq:calc_maxrf}~\cite{Koutini2019Receptive}:

\begin{align}
\label{eq:calc_maxrf}
\nonumber S_n=S_{n-1}*s_n \\
RF_n= RF_{n-1}+(k_n-1)*S_n
\end{align}
where $S_n$, $RF_n$ are cumulative stride and RF of a neuron in the layer $n$ to the  input spectograms. $s_n$, $k_n$ are stride and filter size of layer $n$, respectively. We define the RF of a CNN as the RF of the last convolutional layer.

Equation~\ref{eq:calc_maxrf} shows that we can control the RF of a CNN by changing the filter sizes $k$ or the strides $s$ of different layers or by changing the total number of layers. Koutini et al.~\cite{Koutini2019Receptive} propose a method to gradually decrease the RF of a CNN by changing the filters sizes of deeper convolutional layers to $1 \times 1$, because layers with $1 \times 1$ filters  do not increase the total RF of the CNN. They also show that the effect of RF-regularization is independent on the number of parameters of a network.


We base our investigation on their proposed ResNet~\cite{Koutini2019Receptive,koutinifaresnet2019}, furthermore we adapt the method for a VGG~\cite{simonyanVeryDeepConvolutional2014} based architecture for comparison. We present the details of the architectures in Section~\ref{sec:architectures}.

\subsection{Shake-Shake Regularization}
\label{sec:reg:shake}
Gastaldi~\cite{gastaldi2017shake} proposed Shake-Shake as regularization technique to counter overfitting in computer vision. Shake-Shake works by replacing the summation of the branches of a multi-branch CNN (such as ResNet~\cite{heDeepResidualLearning2016}) with a stochastic affine combination. In other words, the branches' outputs are weighted with random weights that add up to 1, and then summed up (as illustrated in Figure~\ref{fig:shakeshake}).  He shows that  applying Shake-Shake improves the generalization of ResNets on different datasets in computer vision. He proposes a ResNet with three parallel branches  -- two convolution and the identity branch -- where he weights the branches with random weights in forward pass and different random weights in the backward pass. The introduced gradient noise was shown to help in decorrelation  of the branches and therefore countering overfitting. Gastaldi~\cite{gastaldi2017shake} shows  that using the Shake operation, the models achieves better overall performance and better decorrelation between the branches compared to simply summing up the branches. Gastaldi~\cite{gastaldi2017shake} also shows that Shake-Shake networks using   fewer  parameters outperform similar networks on CIFAR-10 and CIFAR-100~\cite{krizhevsky2009cifar}.

%% file: s3_tasks.tex
For the demonstration and analysis of the previously described regularization methods, we choose two specific MIR tasks -- emotion and theme detection and instrument recognition -- along with recently released benchmark datasets. Both are essentials tasks in MIR with various downstream applications such as music browsing, discovery, and recommendation.

In the following we describe both tasks, chosen datasets and recent approaches for modelling them.


\subsection{Emotion and Theme Detection in Music}
\label{sec:mtg:jemando}

The task of emotion and theme detection is a multi-label auto-tagging task where the tags are emotion/mood/theme descriptors of the associated musical piece. It adds to conventional emotion recognition, by using a more diverse set of tags including those that are associated with the mood or theme of the piece. The data comes from a subset of the MTG-Jamendo dataset~\cite{bogdanov2019mtg} released in the Emotion and Theme Recognition in Music Task at the \textit{MediaEval-2019 Benchmark}~\cite{Bogdanov2019mediaeval}. 
The subset contains the samples that are annotated with mood and theme annotations. We will refer to this subset as MTG-Jamendo dataset throughout this paper. The data was collected from the publicly available music collection of Jamendo and the tags are  provided by the content uploaders. 

The dataset has raw audio of full-length songs and the associated tag labels. There are 57 mood/theme labels - some examples being \textit{happy, dark, epic, melodic, love, film, space} etc. 

CNNs dominated the approaches on this dataset at the MediaEval-2019 Benchmark
~\cite{bogdanov2019mtg,koutini2019emotion,sukhavasi2019music_cnnselfattention,amiriparian2019emotion,mediaeval19_inn}. 
The baseline model uses a VGG-like network with five convolutional layers with max-pooling and a single dense layer with dropout. Sukhavasi and Adapa~\cite{sukhavasi2019music_cnnselfattention} use MobileNetV2~\cite{sandler2018mobilenetv2} with self-attention~\cite{vaswani2017attention} to capture temporal relations.
Amiriparian et al~\cite{amiriparian2019emotion} use pre-trained CNNs on Audioset~\cite{audioset2017Gemmeke} and ImagesNet~\cite{dengImagenetLargescaleHierarchical2009} to extract features and recurrent neural networks that operate on these features. 

\subsection{Instrument Recognition}
\label{sec:openmic}
We refer to instrument recognition (also called  instrument identification, or instrument tagging) as the task of detecting the presence or absence of instrument classes in music recordings.  From a machine learning perspective, instrument detection, in its general form, is considered a multi-class, multi-label classification task.

The MIR literature~\cite{GururaniSL19openmoc_attn,GururaniSL18instrument,HanKL17Instrument,Lostanlen2018instrument} distinguishes between three settings for instrument identification, i.e., instrument identification in isolated note recordings, in single instrument sounds, and in recordings which allow simultaneously playing instruments. The third of the three, polyphonic instrument identification, is considered more challenging due to its broader setting, but also more interesting in real-world applications.

In our experiments, we work with OpenMIC-2018~\cite{humphrey2018openmic}, a freely available dataset of $20,000$ music clips for polyphonic instrument classification. The corpus is composed of CC-by licensed recordings collected from the Free Music Archive~\cite{Defferrard2017fda}. Each audio clip has a duration of roughly 10 seconds and is weakly annotated with soft labels. Individual samples are, on average, annotated for roughly two out of twenty instruments, the remaining labels are marked as unknown. For further statistics, and details regarding data collection and annotation, we refer the reader to the original paper~\cite{humphrey2018openmic}.


We are comparing our approach to a set of state-of-the-art methods for instrument tagging on the OpenMIC-2018 dataset. Several of these approaches~\cite{amir2020openmic,GururaniSL19openmoc_attn,humphrey2018openmic} use a pre-trained feature extraction  VGG-like~\cite{simonyanVeryDeepConvolutional2014} CNN trained on AudioSet~\cite{audioset2017Gemmeke}, which generates embeddings of 0.96-second spectrogram snippets to 128-dimensional vectors. These embeddings -- commonly called VGGish embeddings -- are used as inputs for tagging models of varying complexity.
Humphrey et al.~\cite{humphrey2018openmic} use a random-forest model on those features.
Gururani et al.~\cite{GururaniSL19openmoc_attn} compared a fully connected network, a recurrent neural network, and an attention-based model, all trained on the VGGish embeddings. 
They reported performance improvements when incorporating the attention mechanism. 
Anhari~\cite{amir2020openmic} trained a Bi-directional~\cite{hochreiter1997lstm} network combined with an attention mechanism~\cite{KongAttention} on the VGGish embeddings. Castel-Branco et al.~\cite{Branco2020openmic} use a different approach, they relabel clips of OpenMic using external data and report improvements over the baseline.

%% file: s4_archs.tex
In this section, we explain the architectures used in our experiments and how we applied the regularization methods discussed in Section~\ref{sec:regularizers}. We use the same architectures for both tasks and datasets. We vary the receptive field size of each architecture (as explained in their respective sections) and report the results 


\subsection{ResNet}

We use the ResNet~\cite{heDeepResidualLearning2016} proposed in~\cite{Koutini2019Receptive}  as a base architecture and follow their proposed method to apply RF regularization. 
The details of the architecture ( Referred to as  CP\_ResNet) are explained in ~\cite{koutinifaresnet2019}) where a hyper-parameter $\rho$ is introduced to control the receptive field of the network as Explained in Table~\ref{tab_resnet_configs} and Equation ~\eqref{eqn:rfset}.
\begin{equation}
  \label{eqn:rfset}
 x_k = 
     \begin{cases}
        3 &\quad\text{if }k\le \rho \\
       1 &\quad\text{if } k > \rho \\
     \end{cases}
\end{equation}
Changing the $\rho$ value will affect the number of $3 \times 3 $ convolutional layers in the network, and therefore the final receptive field. For example, in order to have a receptive field of  $135\times 135$, we set $\rho=7$. This will result in a CNN configured as explained in Table~\ref{tab_resnet_configs} with $x_k=3 \text{ for }k \in [1,5]$ and  $x_k=1 \text{ otherwise}$. Table~\ref{tab_rho_phi} shows the receptive field of the network to each $\rho$ values.\footnote{ The source code of CP\_ResNet can be found at  \url{https://github.com/kkoutini/cpjku_dcase19/}}
 
This setup allows us to test similar  ResNet variants with gradually varying RF size over the input and study how small changes in the RF size reflects on the networks generalization.

\begin{table}[t]
\caption{The outline of  CP\_ResNet architectures~\cite{koutinifaresnet2019}}
\begin{center}
\begin{tabular}{|c|c|}
\hline
\textbf{RB Number}&\textbf{RB Config} \\
\hline
&Input $ 5 \times 5$ stride=$2$ 
\\
\hline

1&$3 \times 3$, $ 1 \times 1$, P\\
2  & $ x_1 \times x_1$,  $ x_2 \times x_2$, P  \\
3  & $ x_3 \times x_3$,  $ x_4 \times x_4$  \\
4    & $ x_5 \times x_5$,  $ x_6 \times x_6$, P   \\
5&$x_7 \times x_7$, $ x_8 \times x_8$  \\
6 &$ x_9 \times x_9$, $ x_{10} \times x_{10}$  \\
7 &$ x_{11} \times x_{11}$, $ x_{12} \times x_{12}$ \\
8  &$ x_{13} \times x_{13}$, $ x_{14} \times x_{14}$  \\
9 &$ x_{15} \times x_{15}$, $ x_{16} \times x_{16}$  \\
10&$ x_{17} \times x_{17}$, $ x_{18} \times x_{18}$  \\
11 &$ x_{19} \times x_{19}$, $ x_{20} \times x_{20}$  \\
12  &$ x_{21} \times x_{21}$, $ x_{22} \times x_{22}$  \\
\hline
\multicolumn{2}{l}{RB: Residual Block, P: $ 2 \times 2$ max pooling after the block.}\\
\multicolumn{2}{l}{$x_k \in \{ 1 , 3 \}$: hyper parameter we use to control the RF}\\
\multicolumn{2}{l}{ of the network. Number of channels per RB:}\\
\multicolumn{2}{l}{128 for RBs 1-4; 256 for RBs 5-8; 512 for RBs 9-12.} 
\end{tabular}
\label{tab_resnet_configs}
\end{center}
\end{table}

\begin{table}[t]
\caption{Mapping $\rho$ values to the maximum RF of  CP\_ResNet 
(networks configured as in Table~\ref{tab_resnet_configs}).
$\rho$ controls the maximum RF by setting the $x_k$ as explained in Eq.~\eqref{eqn:rfset} ~\cite{koutinifaresnet2019}.}
\begin{center}
\begin{tabular}{|c|c||c|c|}
\hline
\textbf{ $\rho$ value}&\textbf{Max RF}&\textbf{ $\rho$ value}&\textbf{Max RF} \\ \hline
0 & $ 23  \times  23  $
&
1 & $ 31  \times  31  $
\\ \hline
2 & $ 39  \times  39  $
&
3 & $ 55  \times  55  $
\\ \hline
4 & $ 71  \times  71  $
&
5 & $ 87  \times  87  $
\\ \hline
6 & $ 103  \times  103  $
&
7 & $ 135  \times  135  $
\\ \hline
8 & $ 167  \times  167  $
&
9 & $ 199  \times  199  $
\\ \hline
10 & $ 231  \times  231  $
&
11 & $ 263  \times  263  $
\\ \hline
12 & $ 295  \times  295  $
&
13 & $ 327  \times  327  $
\\ \hline
14 & $ 359  \times  359  $
&
15 & $ 391  \times  391  $
\\ \hline
16 & $ 423  \times  423  $
&
17 & $ 455  \times  455  $
\\ \hline
18 & $ 487  \times  487  $
&
19 & $ 519  \times  519  $
\\ \hline
20 & $ 551  \times  551  $
&
21 & $ 583  \times  583  $
\\ \hline
\hline
\end{tabular}
\label{tab_rho_phi}
\end{center}
\end{table}



\subsection{VGG}
In order to compare the performance of ResNet models with VGG-based~\cite{simonyanVeryDeepConvolutional2014} models in different RF configurations we design a VGG based CNN similar to the ResNet described in the previous section. Since VGG~\cite{simonyanVeryDeepConvolutional2014} lacks the residual connection and therefore 
fitting deeper VGG based CNNs can be harder (As shown in ~\cite{heDeepResidualLearning2016} 
), we don't allow the trailing $ 1 \times 1$ convolutional layers. In other words, we iteratively remove $3 \times 3$ convoluational layers instead of replacing them with $ 1 \times 1$ filter. The proposed \emph{base} VGG architecture has the same structure as outlined in Table~\ref{tab_resnet_configs} without the residual connections and has a RF size of $583 \times 583 $.  
We then train the network after removing the last convolutional layer, which decreases the RF of the CNN. We repeat this process for all the previous convolutional layers and report the results.

\subsection{Shake-Shake ResNet}
\begin{figure}[h]
\centering
\includegraphics[]{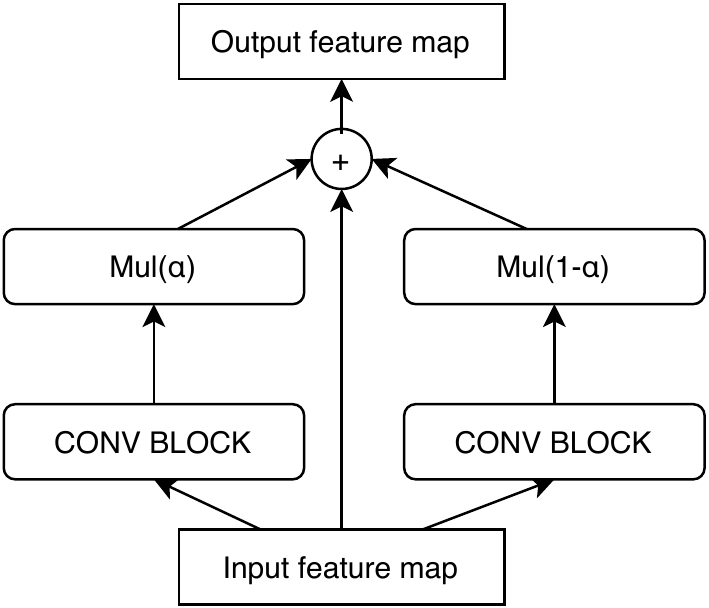}
\caption{Shake-Shake Convolutional block in SS ResNet. $\alpha$ controls the weights of the branches.   }
\label{fig:shakeshake}
\end{figure}

We also design Shake-Shake ResNet (SS ResNet) to have identical RF configurations as the previous CNNs, in order to to be able to compare the different architectures on the same RF size. The base SS ResNet has a similar structure to ResNet  as outlined in Table~\ref{tab_resnet_configs} but with replacing each block (line in the table) with two parallel identical blocks and applying the Shake-Shake operation on their output as explained in Section~\ref{sec:reg:shake} and illustrated in Figure~\ref{fig:shakeshake}. Where $\alpha$ is sampled randomly per input sample in each layer and in both the forward and the backward pass. We set $\alpha=0.5$ when evaluating the models or predicting, which gives equal weights to the branches.


%% file: s5_setup.tex
\subsection{Spectrograms}
We use perceptually-weighted Mel-scaled spectograms with 256  Mel frequency bins and use the training set mean and standard deviation to normalize the CNNs input.
In OpenMIC, where the audio clips have a 10-second length, we use a window size of 2048 and 75\% overlap to extract  a frame in the spectogram with Short Time Fourier
Transform (STFT). However, in the MTG-Jamando datasets the audio clips are longer. Therefore, we randomly sample a smaller clip from the songs in each batch due to memory limitations. We decrease the spectogram extraction window overlap  to 25\%. This increases the time-span covered by two consecutive frames in the spectograms and leading to cover a larger portion of the songs  in a single batch of the CNN input.

\subsection{Model Training}

We use Adam~\cite{kingmaAdamMethodStochastic2014} to train our models. We train all the networks for 150 epochs and report the mean and the standard deviation of the evaluation results of the network in the last 10 epochs. 

We use Mix-up~\cite{zhangMixupEmpiricalRisk2017}, a simple augmentation method that has shown to have great impact on performance and generalization, in  MTG-Jamendo experiments as it showed to increase the performance in our preliminary experiments.  
In contrast, Mix-up did not show performance gains in our preliminary experiments on OpenMIC dataset, therefore the experiments on OpenMIC do not use Mix-up.

For evaluation we use Precision-Recall Area Under Curve (PR-AUC)\footnote{ it's also called ''average precision``} as a metric on both datasets. 
PR-AUC is used since it takes the predictions probability as an input and takes into account all the possible prediction thresholds. We calculate PR-AUC per class and report the macro average over all the classes/tags. This metric was also used in the MediaEval benchmark which allows us to compare with the baseline results and the other approaches.

We additionally report the F-score on OpenMIC to compare with the published approaches on this dataset. It's worth noting that Gururani et al.~\cite{GururaniSL19openmoc_attn} report the average F-score between the positive and negative class of each instrument and then averaged for all instruments. While in~\cite{amir2020openmic} and ~\cite{humphrey2018openmic}\footnote{\url{https://github.com/cosmir/openmic-2018/blob/master/examples/modeling-baseline.ipynb}}, the F-score is calculated per class and averaged for all instruments. We will report both.




%% file: s6_results.tex
\begin{figure}[t]
\centering
\includegraphics[width=3.5in]{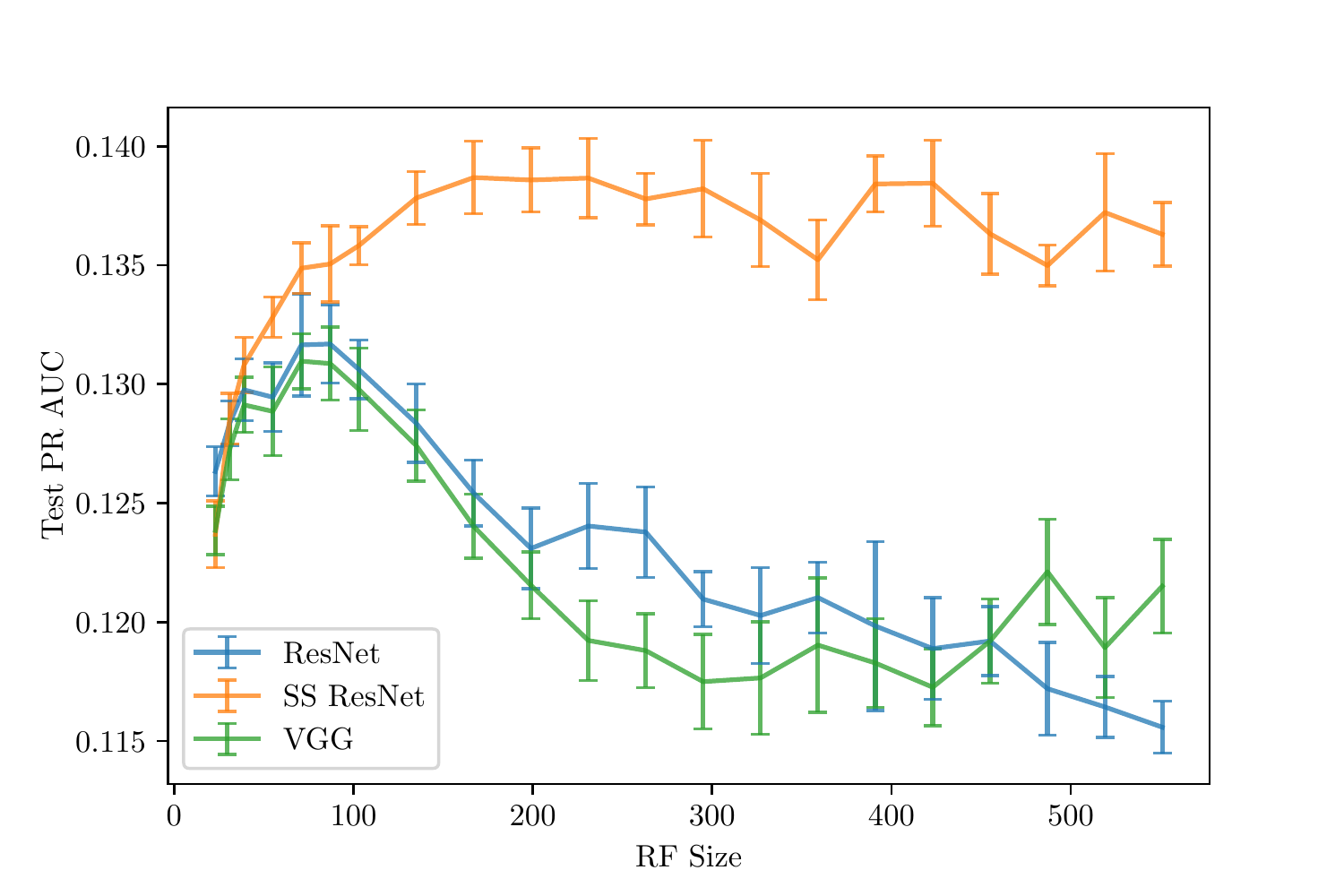}
\caption{Testing PR-AUC of the CNNs  on  MTG-Jamendo   }
\label{fig:mediaeval:main:acc}
\end{figure}

\begin{figure*}[h]
\centering
    \begin{subfigure}[t]{0.5\textwidth}
    \centering
    \includegraphics[width=3.5in]{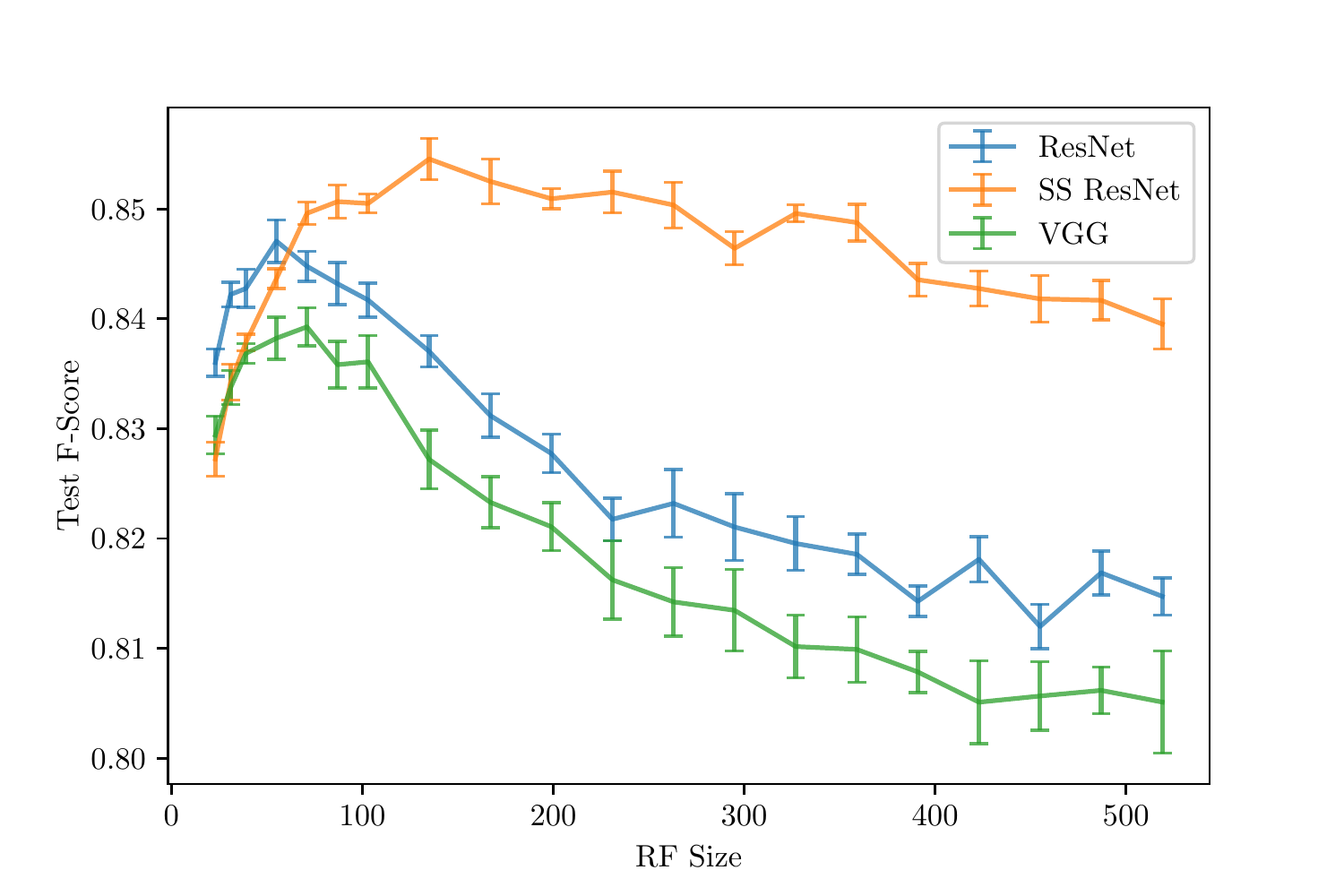}
    \caption{Testing F-Score of CNNs  on OpenMIC dataset.}
    \label{fig:openmic:main:fscore}
    \end{subfigure}%
    ~ 
    \begin{subfigure}[t]{0.5\textwidth}
\centering
\includegraphics[width=3.5in]{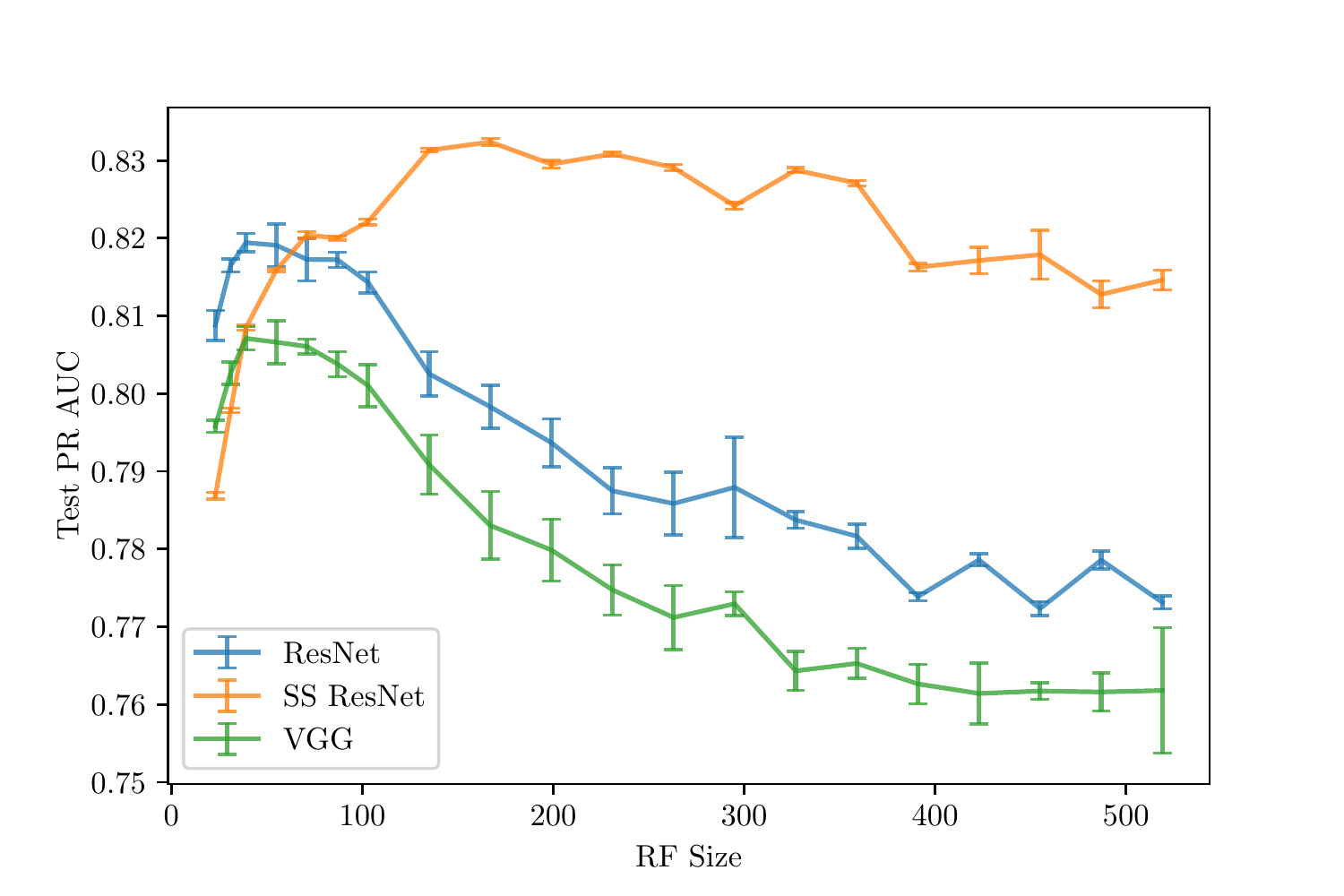}
\caption{Testing PR-AUC of CNNs  on OpenMIC dataset.}
\label{fig:openmic:main:prauc}
\end{subfigure}
\caption{Evaluation Results of CNNs  on OpenMIC dataset.}

\label{fig:openmic:metrics}
\end{figure*}

\subsection{Emotion and Theme Detection in Music}

Figure~\ref{fig:mediaeval:main:acc} shows the PR-AUC  for the studied CNNs (Section~\ref{sec:architectures})  on  MTG-Jamendo Dataset (Section~\ref{sec:mtg:jemando}) for different RF configurations.  The figure shows that the optimal RF size for ResNet and VGG is around $100 $ pixels.  Networks with smaller or larger RF size degrade in performance quickly. However, the figure shows that Shake-Shake regularization diminishes the performance loss in the absence of RF-regularization on this dataset.



Table~\ref{tab:res:mediaEval:sota} shows the comparison of the results of the state-of-the-art  approaches Emotion and Theme Recognition in Music approaches as described in Section~\ref{sec:mtg:jemando}. We compare with single models.  
The table shows that all RF regularized CNNs  achieve a higher performance without pre-training or stacking more complex architectures on the CNNs output.

\begin{table}[h]

\begin{center}
\begin{tabular}{ll}
Method            &  PR\_AUC     \\
\hline
SS ResNet$_{RF=167}$  &     .1387$ \pm.0015 $        \\
ResNet$_{RF=87}$ &      .1317$ \pm  .0016 $    \\
VGG$_{RF=71}$ &      .1309$ \pm  .0011 $    \\
Sukhavasi and Adapa~\cite{sukhavasi2019music_cnnselfattention}  &    .1259           \\
Amiriparian et al~\cite{amiriparian2019emotion}   &    .1175              \\
VGG baseline~\cite{bogdanov2019mtg}  &      .1077                 \\
\hline
\end{tabular}

\end{center}
\caption{
Performance comparison of  the state-of-the-art on the test set of MTG-Jamendo Dataset}
\label{tab:res:mediaEval:sota}
\end{table}

\subsection{Instrument Recognition}

Figures~\ref{fig:openmic:main:fscore} and~\ref{fig:openmic:main:prauc} show F-score and PR-AUC on the OpenMIC dataset (see Section~\ref{sec:openmic}). The figures show that the optimal RF for ResNet and VGG is is approximately between $40 \time 40 $ pixels and  $100 \time 100 $. In addition, their performance degrades rapidly as the RF size of the networks grows or shrinks beyond this range. Similar to the results on the MTG-Jemando Dataset, using Shake-Shake diminishes the negative effect of increasing the RF of the CNNs.

Table~\ref{tab:res:openmic:sota} shows the comparison of the state-of-the-art approaches for  instrument tagging in polyphonic music  approaches on the OpenMIC dataset. For a description of the approaches see Section~\ref{sec:openmic}.
We show that RF-regularized ResNet achieves comparable performance without using external larger datasets or stacking additional models. While by using both RF and Shake-Shake regularization we achieve a better performance than the currently published work.


\begin{table}[h]
\begin{threeparttable}
\centering
\tabcolsep=0.11cm
\begin{tabular}{llll}
Method            &  PR\_AUC   & F-score\tnote{b} & F-score\tnote{a}  \\
\hline
VGG$_{RF=71}$ &      .806$ \pm.002 $    &  .801$ \pm.002 $  &   .839$ \pm.002 $  \\
ResNet$_{RF=55}$ &      .819$ \pm.001 $    &  .809$ \pm.003 $  &   .847$ \pm.002 $  \\
SS ResNet$_{RF=135}$ &      .831$ \pm.000 $ &    .822$ \pm.001$   &  .855$ \pm  .002 $ \\
Castel-Branco ~\cite{Branco2020openmic} &   .701         	            &    -    & -  \\
 Anhari~\cite{amir2020openmic} &   -        	            &    -    &  .83\tnote{c}  \\

Gururani~\cite{GururaniSL19openmoc_attn}      &    -    &  .81\tnote{c} & - \\
 Baseline~\cite{humphrey2018openmic} &      .795  &   .785  &  .826 \\
\hline
\end{tabular}
\begin{tablenotes}
\item[a] Classical F-score as used in~\cite{amir2020openmic,humphrey2018openmic}
\item[b]Average of the F-score of the positive and negative class  per instrument as proposed in  ~\cite{GururaniSL19openmoc_attn} 
\item[c] Indicates that the number is approximately read from the figures in the referenced papers
\end{tablenotes}
\end{threeparttable}
\caption{Performance  Comparison of  the state-of-the-art on the test set of OpenMIC. }
\label{tab:res:openmic:sota}
\end{table}

\begin{figure*}[h]
\centering
     \begin{subfigure}[t]{0.5\textwidth}
        \centering
        \includegraphics[width=3.5in]{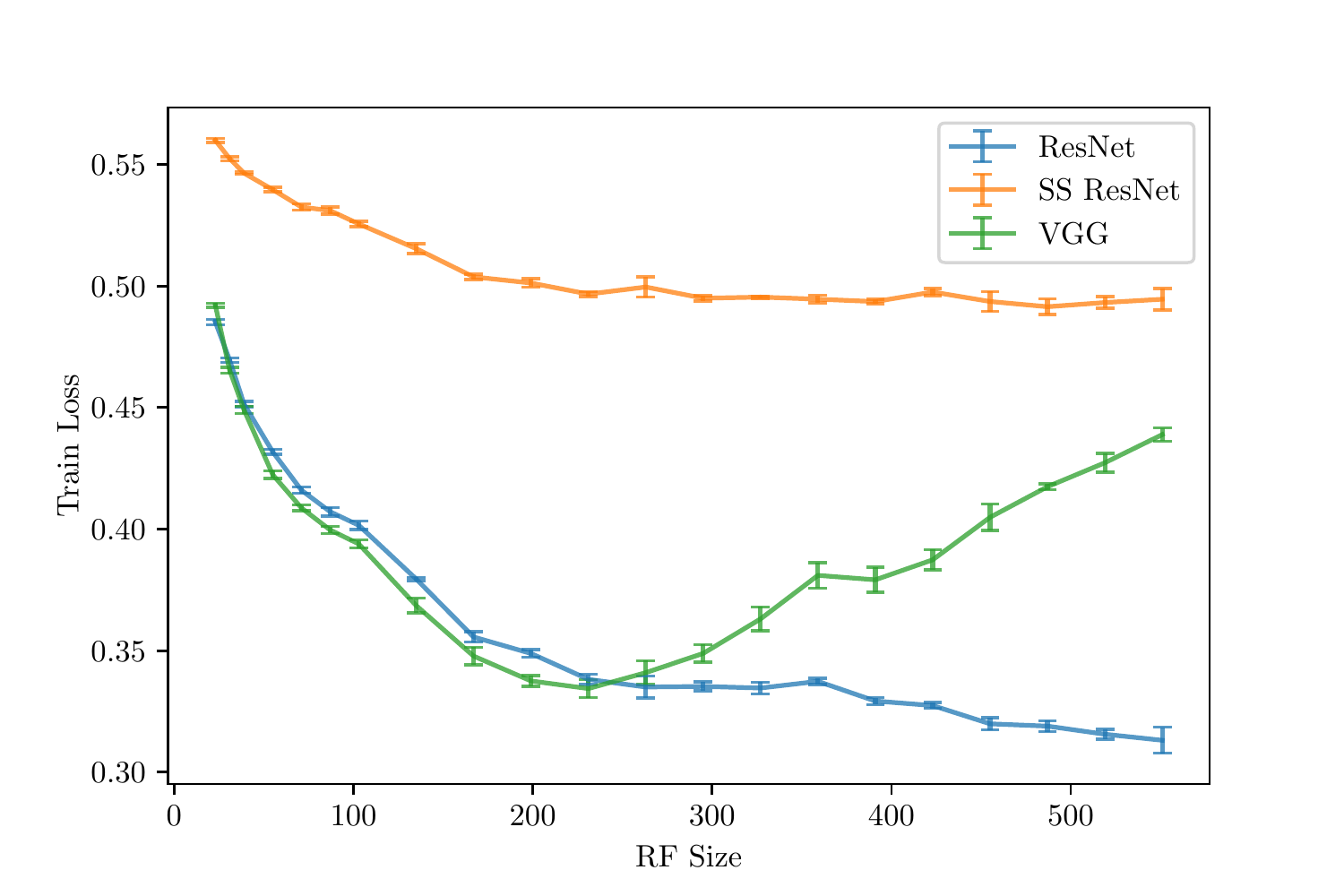}
        \caption{Training loss of the CNNs  on  MTG-Jamendo   }
        \label{fig:mediaeval:train:loss}

    \end{subfigure}%
    ~ 
    \begin{subfigure}[t]{0.5\textwidth}
        \centering
    \includegraphics[width=3.5in]{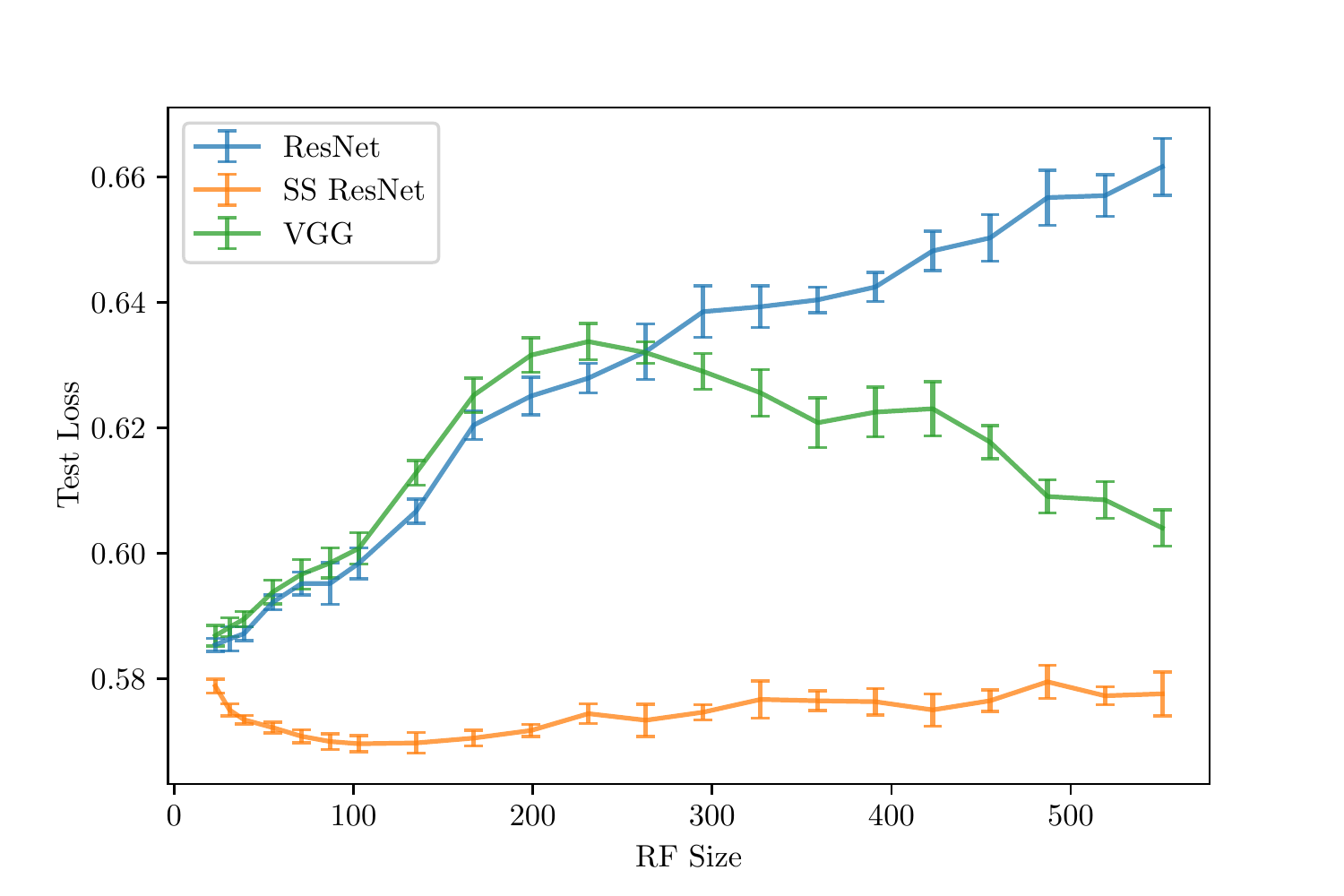}
    \caption{Testing loss of the CNNs  on  MTG-Jamendo   }
    \label{fig:mediaeval:test:loss}
    \end{subfigure}

        \begin{subfigure}[t]{0.5\textwidth}
    \centering
    \includegraphics[width=3.5in]{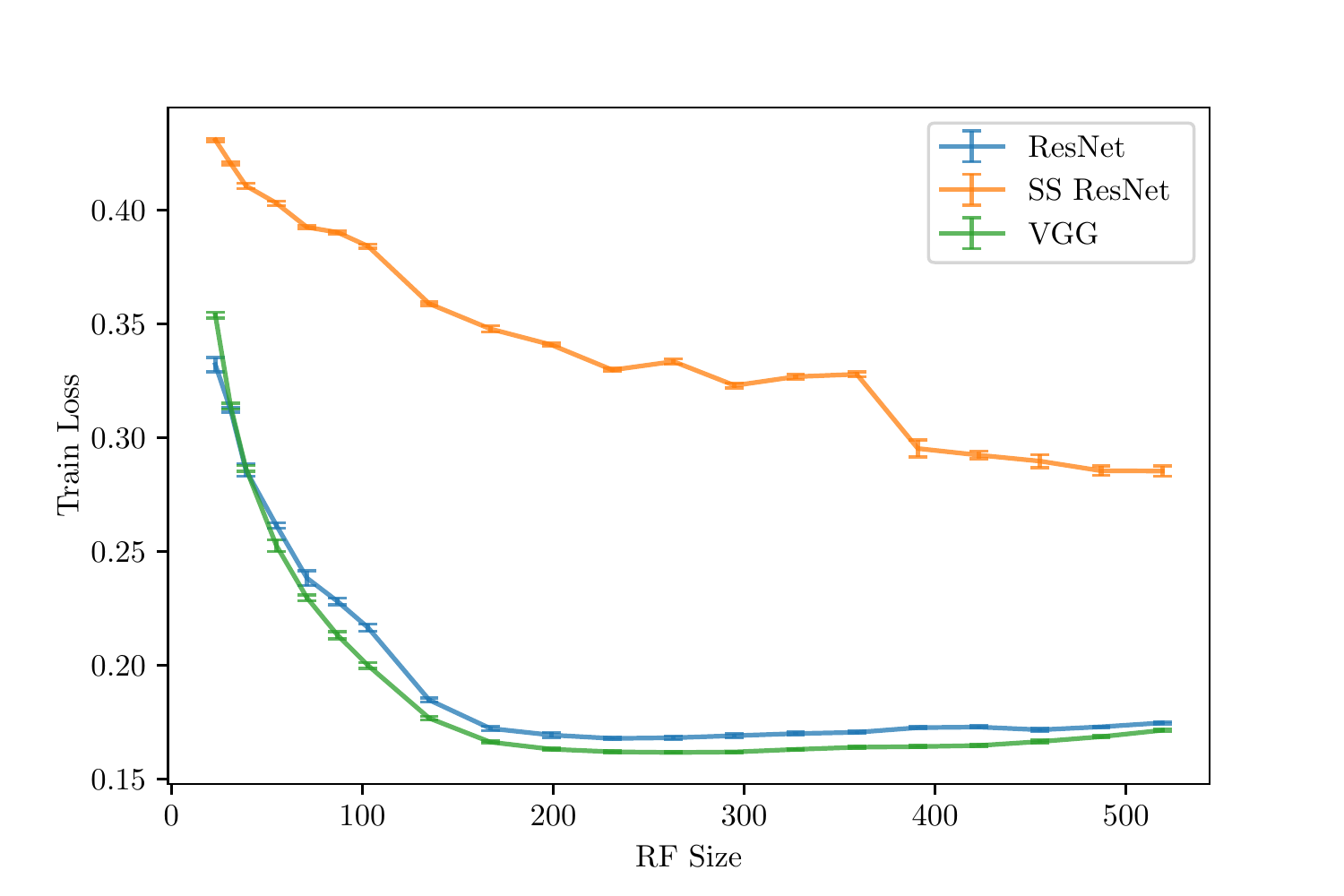}
    \caption{Training Loss of CNNs  on OpenMIC dataset.}
    \label{fig:openmic:train:loss}
    \end{subfigure}%
~
    \begin{subfigure}[t]{0.5\textwidth}
\centering
\includegraphics[width=3.5in]{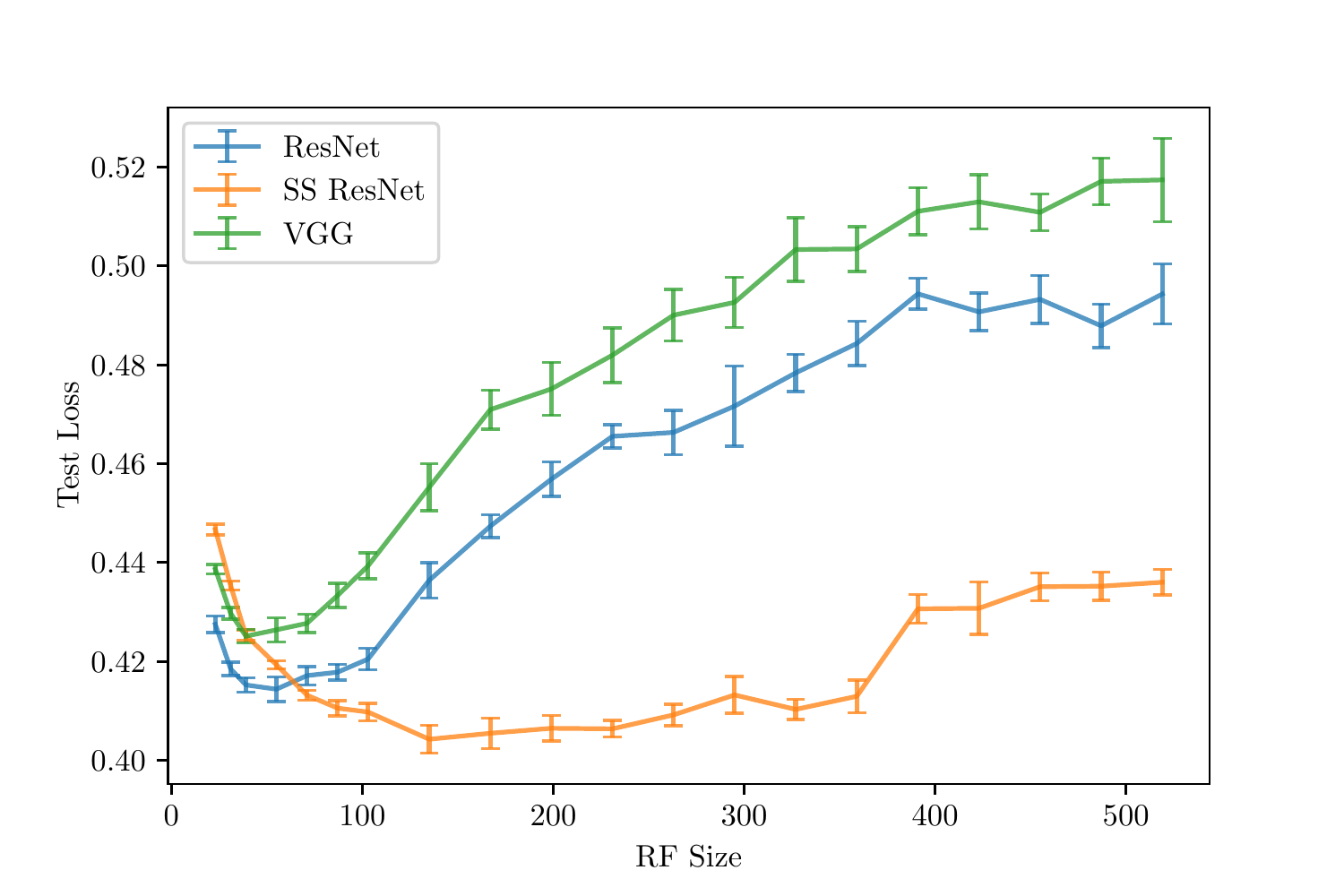}
\caption{Testing Loss of CNNs  on OpenMIC dataset.}
\label{fig:openmic:test:loss}
\end{subfigure}
    \caption{Comparing the training and the testing loss of different CNNs }
\label{fig:compare:loss}
\end{figure*}

\subsection{Generalization Analysis}

Figures~\ref{fig:mediaeval:main:acc} and~\ref{fig:openmic:metrics} show that ResNet tends to outperform VGG when both architectures have the same RF size across most RF configurations, in a way that agrees their performance in the vision domain. The figures  show that using RF-regularization is crucial for the architectures performance and generalization on the unseen test sets, especially in the absence of strong regularization such as Shake-Shake. These results are in accordance with the findings of Koutini et al.~\cite{Koutini2019Receptive} on acoustic scene classification datasets. 

We compare the training loss with testing loss in Figure~\ref{fig:compare:loss} for the studied CNNs over different RF size configurations on both datasets. As a rule, we see that that the training loss decreases with the increase of the RF size, which indicates increasing the CNN ability to fit the training data. However, this does not translate and generalize to the testing data as seen by the testing loss increase when the RF size goes beyond an optimal range. The figure also shows that Shake-Shake reduces the testing loss increase w.r.t RF size increase as it hinders overfitting the training data. As a side effect, however, using the Shake-Shake regularization, SS ResNet requires a larger RF than the optimal RF-size range of ResNet and VGG on these datasets. 

We observe that the training loss of the VGG architecture increases on larger RF sizes on the MTG-Jemando dataset. We believe this is due to the architecture design (explained in Section~\ref{sec:architectures} ). In other words, the RF increases with deeper VGG networks, and deeper VGGs suffer from the vanishing gradient  and this partially prevent the network from fitting the training data. Remarkably, this translated to a slight testing performance increase on this dataset. However, we don't see the same effect on the OpenMIC.

%% file: s7_conclusion.tex
In this paper, we analysed two regularization methods for CNN architectures, namely RF regularization and Shake-Shake regularization.
Our results suggests that common deep unregularized CNNs -- which, as we showed, have a larger RF -- tend to overfit on the training data and therefore perform poorly on test data.
We showed that RF-regularization is crucial for improving the generalisation of such architectures on the musical tasks.
Additionally, we demonstrated the potential of Shake-Shake regularization on the investigated tasks,
and reducing the overfitting. 
Finally, we showed that by combining the two regularizations, we can achieve state-of-the-art performance, 
without using pre-trained models on larger datasets, or using complex modules such as attention.
Based on these results, we offer our regularized CNNs as a new baseline for the investigated datasets.
